\documentstyle[epsfig,aps]{revtex}
\begin{document}
\draft
\title{Stability of Transparent Spherically Symmetric Thin Shells and Wormholes}
\author{Mustapha Ishak\cite{email} and Kayll Lake\cite{email1}}
\address{Department of Physics, Queen's University, Kingston, Ontario, Canada, K7L 3N6 }
\date{\today}
\maketitle
\date{\today}
\maketitle
\begin{abstract}
The stability of transparent spherically symmetric thin shells
(and wormholes) to linearized spherically symmetric perturbations
about static equilibrium is examined. This work generalizes and
systematizes previous studies and explores the consequences of
including the cosmological constant. The approach shows how the
existence (or not) of a domain wall dominates the landscape of
possible equilibrium configurations.
\end{abstract}
\pacs{04.20.Cv,04.20.Gz,04.70.Bw}
\section{Introduction}
It is difficult to imagine a geometrical construction more
fundamental than an embedded hypersurface across which the second
fundamental form is discontinuous. Studies of such constructions
within the Darmois \cite{Darmois} - Israel \cite{Israel} formalism
(and others) have seen widespread application in classical general
relativity (e.g. \cite{mus}). The importance of such a
construction is of course not limited to classical general
relativity but is in a real sense central to a much wider class of
considerations, such as brane world cosmologies \cite{xxx}.

In this paper we examine the stability of spherically symmetric
thin shells, which are in a clearly defined sense `` transparent",
to linearized spherically symmetric perturbations about static
equilibrium under the assumption that the shells remain
transparent under perturbation. The study of the stability of
shells is not new \cite{chase}. Here we follow Poisson and Visser
\cite{poissonandvisser} and parameterize the stability of
equilibrium so that we do not have to specify a surface equation
of state. The present work generalizes and systematizes previous
studies and shows how the existence (or not) of a domain wall
\cite{domainwall} dominates the landscape of possible equilibrium
configurations.
\section{Stability}
\subsection{Equation of Motion}
We use the notation of Musgrave and Lake \cite{mus} throughout and
consider a 3-surface $\Sigma$ dividing spherically symmetric
spacetime into two distinct parts. In the spherically symmetric
case, in terms of the intrinsic coordinates $(\tau,\theta,\phi)$,
the metric intrinsic to $\Sigma$ can be given as
\begin{equation}
ds_{\Sigma}^2 = R^2(\tau)d\Omega^2+\epsilon d\tau^2 \label{intrinsic}
\end{equation}
where $\Sigma$ is timelike (spacelike) for $\epsilon = -1 (+1)$
and $d\Omega^2$ is the metric of a unit sphere. In terms of the
surface energy density $\sigma(\tau)$, the surface mass of
$\Sigma$ is defined by
\begin{equation}
M(\tau) = 4 \pi R^2 \sigma = \epsilon \gamma_{\theta \theta} \label{surfacemass}
\end{equation}
where $\gamma_{\theta \theta} = [K_{\theta \theta}]$  and $K_{ab}$
denotes the second fundamental form of $\Sigma$. (Our notation is
explained in \cite{notation}.) We have
\begin{equation}
K_{\theta \theta}^2 = R^2\Big(\dot{R}^2-\epsilon(1-\frac{2 m}{R})\Big) \label{kthetatheta}
\end{equation}
where $m$ (in general not a constant) is the (invariantly defined
\cite{mass}) effective gravitational mass of one of the enveloping
spacetimes. The identity
\begin{equation}
K_{\theta \theta}^{+ 2} =  \frac{1}{4M^{2}}(K_{\theta \theta}^{+ 2} - K_{\theta \theta}^{- 2} + M^2)^2 \label{identity}
\end{equation}
can be written in the form \cite{lake}
\begin{equation}
\label{eqn-lake}
\dot{R}^2 = \epsilon + \left( \frac{\left[ m \right]}{M} \right)^2
   - \frac{2 \epsilon \overline{m}}{R} +
   \left( \frac{M}{2R} \right)^2.
\end{equation}
We refer to (\ref{eqn-lake}) as ``the equation of motion" for
$\Sigma$. This is a quadratic identity and applies to both shells
and wormholes.
\subsection{Transparency Condition}
In this paper we use the transparency condition
\begin{equation}
[G^\alpha_\beta n_\alpha u^\beta]=0, \label{transparency}
\end{equation}
where $G^\alpha_\beta$ is the (4-dimensional) Einstein tensor,
$u^\alpha$ is the 4-tangent to $\Sigma$ and $n^\alpha$ the
4-normal to $\Sigma$. It follows from (\ref{transparency}) that
for $\sigma^{'}\neq0$
\begin{equation}
\Big(\frac{M}{2R}\Big)^{''} =\frac{\Upsilon}{2R^3}\Big(1+2\frac{P^{'}}{\sigma^{'}}\Big) \label{transparency1}
\end{equation}
where
\begin{equation}
\Upsilon \equiv 3 M -(MR)^{'} = 8 \pi R^2(\sigma + P),
\label{upsilon}
\end{equation}
$'\equiv \frac{d}{dR}$ and $P$ denotes the intrinsic surface
pressure.
\subsection{Restrictions from the Potential}
In (\ref{eqn-lake}) $m$ is defined in terms of coordinates extrinsic to $\Sigma$. In order to view (\ref{eqn-lake}) as defining the potential $V(R)$, where
\begin{equation}
\dot{R}^2 = -V(R), \label{potential}
\end{equation}
$R$ must define $m$ uniquely. This imposes a restriction on the enveloping spacetimes. Consider, for example, Bondi coordinates:
\begin{equation}
ds^2 = 2c(v,r)dvdr-c^2(v,r)\Big(1-\frac{2m(v,r)}{r}\Big)dv^2+r^2d\Omega^2. \label{bondi}
\end{equation}
The continuity of the intrinsic metric gives
\begin{equation}
r^+ = r^- = R \label{intriniccont}
\end{equation}
so that for (\ref{potential}) to hold at equilibrium we require
\begin{equation}
m=m(r)=m(R) \label{mass}
\end{equation}
which, we note, is precisely the condition that guarantees
(\ref{transparency}) at equilibrium. For (\ref{transparency}) to
hold away from equilibrium it follows that $\frac{\partial
c}{\partial r }=0$.  In the remainder of this work we assume
(\ref{mass}) and $\frac{\partial c}{\partial r }=0$ so that the
shells remain transparent under perturbation. For timelike
$\Sigma$ then our assumptions require static embeddings in the
neighborhood of $\Sigma$.
\subsection{$\Upsilon=0$}
In this paper we consider linearized spherically symmetric
perturbations about static equilibrium. Following Poisson and
Visser \cite{poissonandvisser}, we use
$\frac{P^{'}}{\sigma^{'}}(R)$ as a parameterization of the
stability of equilibrium so that we do not have to specify an
equation of state on $\Sigma$. This parameterization singles out
the case $\Upsilon=0$, equivalently
\begin{equation}
(MR)^{'}=3M,     \qquad      (\frac{M}{2R})^{''}=0, \qquad \sigma
= -P, \label{posupsilonzero}
\end{equation}
where $\sigma^{'}=0$. This case (a ``domain wall"
\cite{domainwall}) signals a fundamental change in the surfaces
that distinguish stable equilibria parameterized by
$\frac{P^{'}}{\sigma^{'}}(R)$ as it enters as an asymptote
($\frac{P^{'}}{\sigma^{'}} \rightarrow \pm \infty$). Whereas this
limit has well known concrete manifestations, for example Casimir
and false vacua (\textit{e.g.} \cite{poissonandvisser}), it
circumvents the usual phenomenology of $\Sigma$ (\textit{e.g.}
\cite{mus}) since the surface stress-energy tensor of $\Sigma$ is
simply proportional to the metric intrinsic to $\Sigma$. In what
follows we consider $\Upsilon\neq0$. Whether or not this asymptote
enters into a given configuration depends simply on the nature of
the roots to the polynomial $\sigma + P=0$. Phenomenologically,
for timelike $\Sigma$, $\frac{P^{'}}{\sigma^{'}}(R)$ represents
the square of the speed of sound on $\Sigma$ and so one would
normally expect $0\leq\frac{P^{'}}{\sigma^{'}}(R)<1$. However, one
could also construct $\Sigma$ out of (say) fields for which
$\frac{P^{'}}{\sigma^{'}}(R)$ need not even be positive. Since we
are concerned with stability, not phenomenology, the plots we give
generally extend beyond $0\leq\frac{P^{'}}{\sigma^{'}}(R)<1$.
\subsection{Equilibrium}
From (\ref{eqn-lake}) and (\ref{potential}) it follows that the
equilibrium condition $V^{'}=0$ is
\begin{equation}
\Big(\frac{M}{2R}\Big)^{'} = \frac{2R}{M}\Big\{\epsilon\Big(\frac{\overline{m}}{R}\Big)^{'}-\frac{[m]}{M}\Big(\frac{[m]}{M}\Big)^{'} \Big\} \equiv \Gamma, \label{vprime}
\end{equation}
and the stability condition $V^{''}>0$ is
\begin{equation}
\frac{M}{2R}\Big(\frac{M}{2R}\Big)^{''} < \Psi-\Gamma^2, \label{vprimeprime}
\end{equation}
where
\begin{equation}
\Psi \equiv \epsilon\Big(\frac{\overline{m}}{R}\Big)^{''}-\frac{[m]}{M}\Big(\frac{[m]}{M}\Big)^{''}-\Big(\frac{[m]}{M}\Big)^{' 2}. \label{Psi}
\end{equation}
With the aide of the transparency condition (\ref{transparency1}) then the conditions for stable equilibria are
\begin{equation}
1+2\frac{P^{'}}{\sigma^{'}}<\Phi,     \qquad      M\Upsilon>0,
\label{posupsilon}
\end{equation}
and
\begin{equation}
1+2\frac{P^{'}}{\sigma^{'}}>\Phi,     \qquad       M\Upsilon<0,
\label{negupsilon1}
\end{equation}
where
\begin{equation}
\Phi \equiv \frac{4R^4}{M \Upsilon}\Big(\Psi-\Gamma^2\Big).
\label{Phidef}
\end{equation}
It is perhaps worthy of note that not only is there no need for
the specification of an equation of state on $\Sigma$, it is not
necessary to even calculate $P$. It is only necessary to calculate
$P$ for a background check on the existence of the asymptote
$\Upsilon=0$ by way of a search for allowed roots to the
polynomial $\sigma +P=0$.
\section{Examples}
In the following examples the explicit form of $\Phi$ can become
lengthy. We find it instructive to demonstrate the possible
equilibrium configurations graphically. In the approach we have
used, the appearance (or not) of the asymptote $\Upsilon=0$
dominates the landscape of possible equilibrium configurations.
The polynomial $\sigma+P=0$ is given in Appendix A.
\subsection{Schwarzschild wormholes}
Poisson and Visser \cite{poissonandvisser} have considered the
stability of Schwarzschild wormholes ($\gamma_{\theta \theta} = 2
K_{\theta \theta}^{+}$). In our notation they consider the case
\begin{equation}
\overline{m} = m, \qquad [m]=0, \qquad \epsilon=-1, \qquad M=-2R\sqrt{1-2\frac{m}{R}} \label{wormhole}
\end{equation}
where $m$ is a constant. It follows that $\Phi$ takes a
particularly simple form:
\begin{equation}
\Phi=\frac{m(3m-2R)}{(R-2m)(R-3m)}.\label{wormhole1}
\end{equation}
Equations (\ref{posupsilon}), (\ref{negupsilon1}) and
(\ref{wormhole1}) reproduce their calculation.  Physical units
associated with this case are given in Appendix B. The situation
is generalized in Figure 1 where we have not assumed the
continuity of $m$.  For this example $M<0$ and $M\Upsilon =
4R(R-3m)$ so that the region \textit{above} the surface
\textit{(a)} and \textit{below} the surface \textit{(b)} in Figure
1 correspond to stable equilibrium.
\subsection{Schwarzschild black holes}
Brady, Louko and Poisson \cite{brady} have considered the
stability of a shell surrounding a classical Schwarzschild black
hole and have shown that stability conditions supersede the energy
conditions  for this configuration as considered previously by
Frauendiener, Hoenselaers and Konrad \cite{frauen}. In this case
\begin{equation}
m^{+}>m^{-}, \qquad \epsilon=-1, \qquad
M=R\sqrt{1-2\frac{m^{-}}{R}}-R\sqrt{1-2\frac{m^{+}}{R}}
\label{blackhole}
\end{equation}
where we have labelled $``+"$ \textit{exterior} to the shell in
the sense that the event horizon is in $``-"$. Since $M$ and
$\Upsilon$ are $>0$ it follows that the region \textit{below} the
surface shown in Figure 2 corresponds to stable equilibrium. We
find that $\frac{R}{m^{+}}\sim 2.37$ as $m^{-}\rightarrow 0$ and
that $\frac{R}{m^{+}}\sim 3$ as $m^{-}\rightarrow m^{+}$ so that
$R>3m^{-}$ in agreement with previous results. There is no
asymptote $\Upsilon=0$ in this case.
\subsection{Cosmological Constant}
In what follows we explore the physical consequences of including
a cosmological constant ($\Lambda$) in the background. The only
spacetimes we consider here are the Schwarzschild -de Sitter and
Schwarzschild - anti de Sitter spaces so that the effective
gravitational mass $m(r)$ is given by
\begin{equation}
m(r) = \texttt{m} + \frac{\Lambda r^3}{6},\label{mlambda}
\end{equation}
where $\texttt{m}$ is a constant.
\subsubsection{Schwarzschild -de Sitter $|$ Schwarzschild}
Qualitatively, the inclusion of a positive cosmological constant
inside a timelike shell should weaken the binding of the shell in
the sense that $P$ will become negative (a surface
\textit{tension}) to stablize the shell at sufficiently large $R$.
The result is quite unlike the case $\Lambda=0$ (above) were
surface tensions are not required to stablize the shell at any
radius. This effect is demonstrated in Figure 3. In this case
\begin{equation}
m^{+}>\texttt{m}^{-}, \qquad \epsilon=-1, \qquad
M=R\sqrt{1-2\frac{\texttt{m}^{-}}{R}-\frac{\Lambda R^2}{3}
}-R\sqrt{1-2\frac{m^{+}}{R}} \label{des1}
\end{equation}
where again we have labelled $``+"$ \textit{exterior} to the shell
in the sense that the innermost event horizon is in $``-"$. Since
$M$ and $\Upsilon$ are $>0$ in this case, it follows that the
region \textit{below} the surface shown in Figure 3 corresponds to
stable equilibrium. There is no asymptote $\Upsilon=0$ in this
case.
\subsubsection{Schwarzschild - anti de Sitter $|$ Schwarzschild}
This case is the same as the previous one except that $\Lambda <
0$. Qualitatively, the inclusion of a negative cosmological
constant inside a timelike shell should not \textit{require} a
surface tension to stablize the shell at sufficiently large $R$.
Whereas this agrees with what we find (see Figure 4), in this case
we find an asymptote $\Upsilon=0$ with the consequence that the
stability surfaces resemble the wormhole case. However, since
$M>0$ here, the regions of stability are the reverse of those for
wormholes in agreement with our intuition.
\subsubsection{Inversions of 1 and 2}
By the inversion of $1$ we mean
\begin{equation}
\texttt{m}^{+}>m^{-}, \qquad \epsilon=-1, \qquad
M=R\sqrt{1-2\frac{m^{-}}{R}
}-R\sqrt{1-2\frac{\texttt{m}^{+}}{R}-\frac{\Lambda R^2}{3}}
\label{des1inv}
\end{equation}
with $\Lambda > 0$ and where again we have labelled $``+"$
\textit{exterior} to the shell in the sense that the innermost
event horizon is in $``-"$. In this case we find that the
stability regions are qualitatively similar to Figure 4. The
inverse of $2$ is the same situation with $\Lambda < 0$ and in
this case we find that the stability regions are qualitatively
similar to Figure 3 in agreement with intuition.
\subsubsection{Schwarzschild - (anti) de Sitter $|$ Schwarzschild - (anti) de Sitter}
One would not expect that the inclusion of $\Lambda$
\textit{continuous} across $\Sigma$ would change, in a significant
way, possible equilibrium configurations. To test this consider
\begin{equation}
\texttt{m}^{+}>\texttt{m}^{-}, \qquad \epsilon=-1, \qquad
M=R\sqrt{1-2\frac{\texttt{m}^{-}}{R}-\frac{\Lambda R^2}{3}
}-R\sqrt{1-2\frac{\texttt{m}^{+}}{R}-\frac{\Lambda R^2}{3}}
\label{des2}
\end{equation}
where again we have labelled $``+"$ \textit{exterior} to the shell
in the sense that the innermost event horizon is in $``-"$. Our
findings are summarized in Figure 5 which, when compared to Figure
2, agrees with what we would expect.
\section{Conclusions}
In this paper we have examined the stability of transparent
spherically symmetric thin shells to linearized spherically
symmetric perturbations about static equilibrium under the
assumption that the shells remain transparent under perturbation.
Our approach has followed that due to Poisson and Visser
\cite{poissonandvisser} wherein $\frac{P^{'}}{\sigma^{'}}(R)$ is
used as a parameterization of the stability of equilibrium. An
equation of state on $\Sigma$ is not specified. The present work
generalizes and systematizes previous studies. We have found that
the single feature which dominates the landscape of equilibrium
configurations is the appearance or not of the asymptote
$\Upsilon=0$. The cosmological constant has been included in our
calculations with results in agreement with intuition. We find
that the closest radius for a stable shell is $R = 3
\texttt{m}^{-}$ irrespective of $\Lambda$ and at that radius
$\frac{P^{'}}{\sigma^{'}}=1$.
\section*{Acknowledgments}
This work was supported by a grant (to KL) from the Natural
Sciences and Engineering Research Council of Canada and by an
Ontario Graduate Scholarship (to MI). Portions of this work were
made possible by use of \textit{GRTensorII}\cite{grt}.

\newpage
\begin{appendix}
\begin{center}
 \textbf{Appendix A: The Asymptote $\Upsilon=0$}
\end{center}
 The existence of the asymptote $\Upsilon=0$ depends on the
 existence of allowed roots to the equation $\sigma + P=0$.
 Writing
 \begin{equation}
 \texttt{f}(R) \equiv 1-\frac{2m(R)}{R} \label{fdef}
 \end{equation}
 these roots are determined by the polynomial
 \begin{equation}
 \delta \sqrt{F}(2f-Rf^{'})=\sqrt{f}(2F-RF^{'}) \label{poly}
 \end{equation}
 where $\delta=+1$ for shells, $\delta=-1$ for wormholes and we
 have distinguished the sides of $\Sigma$ by $\texttt{f}=f$ and $\texttt{f}=F$.
 As an example, the solution to equation (\ref{upsilon}) for the Schwarzschild
 wormhole case (see Figure 1) is shown in Figure 6.

\end{appendix}

\newpage
\begin{appendix}
\begin{center}
 \textbf{Appendix B: Physical Units and Schwarzschild wormholes}
\end{center}
\begin{table}
\begin{tabular}{|c|c|c|c|c|c|c|c|c|}
R(km) & $R/M$ & $M/M_\odot$ & $(P'/\sigma')_{min}$ & & R(km) & $R/M$ & $M/M_\odot$ & $(P'/\sigma')_{max}$\\
1 & -1.0 & -0.67721 & 4.75 &   & 1 & -0.55 & -1.23130 & -0.62\\
1 & -1.5 & -0.45148 & 3.50 &   & 1 & -0.60 & -1.12869 & -0.81\\
1 & -2.0 & -0.33861 & 4.98 &   & 1 & -0.65 & -1.04187 & -1.12\\
1 & -2.5 & -0.27089 & 7.14 &   & 1 & -0.70 & -0.96745 & -1.64\\
1 & -3.0 & -0.22574 & 9.84 &   & 1 & -0.75 & -0.90295 & -2.69\\
1 & -10.0 & -0.06772 & 100.76 &   & 1 & -0.85 & -0.79672 & -25.80\\
5 & -1.0 & -3.38607 & 4.75 &   & 5 & -0.55 & -6.15648 & -0.62\\
5 & -1.5 & -2.25738 & 3.50 &   & 5 & -0.60 & -5.64344 & -0.81\\
5 & -2.0 & -1.69303 & 4.98 &   & 5 & -0.65 & -5.20933 & -1.12\\
5 & -2.5 & -1.35443 & 7.14 &   & 5 & -0.70 & -4.83724 & -1.64\\
5 & -3.0 & -1.12869 & 9.84 &   & 5 & -0.75 & -4.51475 & -2.69\\
5 & -10.0 & -0.33861 & 100.76 &   & 5 & -0.85 & -3.98361 & -25.80\\
10 & -1.0 & -6.77213 & 4.75 &   & 10 & -0.55 & -12.31297 & -0.62\\
10 & -1.5 & -4.51475 & 3.50 &   & 10 & -0.60 & -11.28689 & -0.81\\
10 & -2.0 & -3.38607 & 4.98 &   & 10 & -0.65 & -10.41866 & -1.12\\
10 & -2.5 & -2.70885 & 7.14 &   & 10 & -0.70 & -9.67447 & -1.64\\
10 & -3.0 & -2.25738 & 9.84 &   & 10 & -0.75 & -9.02951 & -2.69\\
10 & -10.0 & -0.67721 & 100.76 &   & 10 & -0.85 & -7.96721 & -25.80\\
100 & -1.0 & -67.72131 & 4.75 &   & 100 & -0.55 & -123.12966 & -0.62\\
100 & -1.5 & -45.14754 & 3.50 &   & 100 & -0.60 & -112.86886 & -0.81\\
100 & -2.0 & -33.86066 & 4.98 &   & 100 & -0.65 & -104.18664 & -1.12\\
100 & -2.5 & -27.08853 & 7.14 &   & 100 & -0.70 & -96.74473 & -1.64\\
100 & -3.0 & -22.57377 & 9.84 &   & 100 & -0.75 & -90.29508 & -2.69\\
100 & -10.0 & -6.77213 & 100.76 &   & 100 & -0.85 & -79.67213 & -25.80\\
\end{tabular}

\caption{Physical units associated with the wormhole case with
continuous mass. Values for the upper branch in Figure 7 are shown
on the left.}
\end{table}

\end{appendix}
\newpage
\begin{figure}
\psfig{file=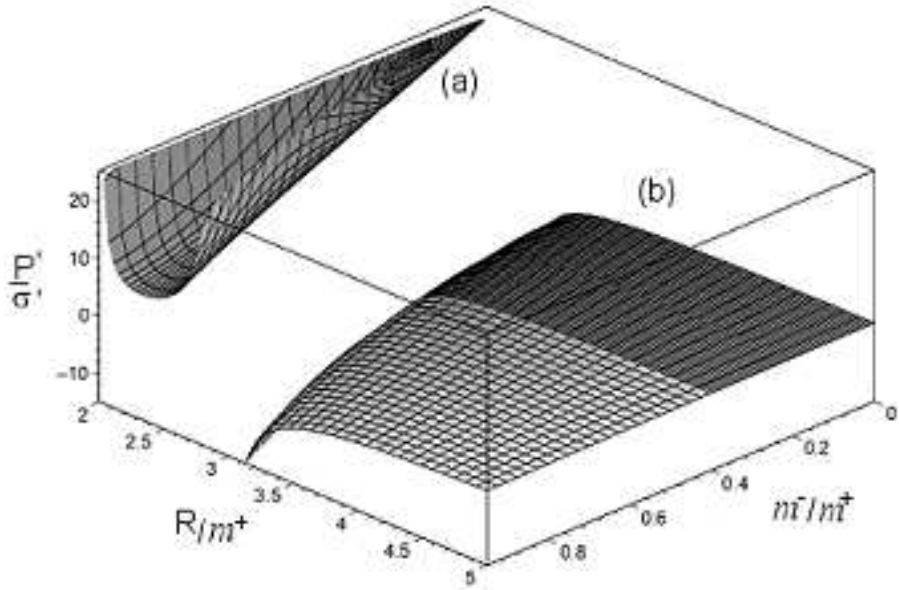}\\ \caption{A generalization of the work by
Poisson and Visser in which $m$ is not assumed continuous across
$\Sigma$. The region \textit{above} the surface \textit{(a)} and
\textit{below} the surface \textit{(b)} correspond to stable
equilibrium. The figure given by Poisson and Visser is the slice
$m^{-}=m^{+}$. The position of the asymptote $\Upsilon=0$ for this
case is shown in Figure 6. } \label{fig1}
\end{figure}

\newpage

\begin{figure}
\psfig{file=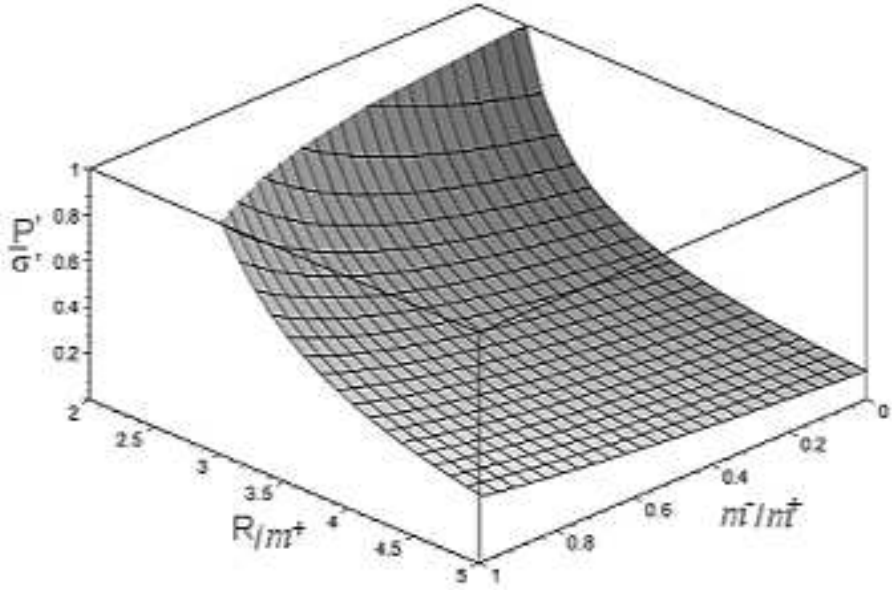}\\ \caption{A reproduction of the classical
Schwarzschild black hole case examined by Brady, Louko and
Poisson. Since $M$ and $\Upsilon>0$ it follows that the region
\textit{below} the surface shown corresponds to stable
equilibrium. For $\frac{P^{'}}{\sigma^{'}}=1$ we find that
$\frac{R}{m^{+}}\sim 2.37$ as $m^{-}\rightarrow 0$ and that
$\frac{R}{m^{+}}\sim 3$ as $m^{-}\rightarrow m^{+}$ so that
$R>3m^{-}$ in agreement with previous results.} \label{fig2}
\end{figure}

\newpage

\begin{figure}
\psfig{file=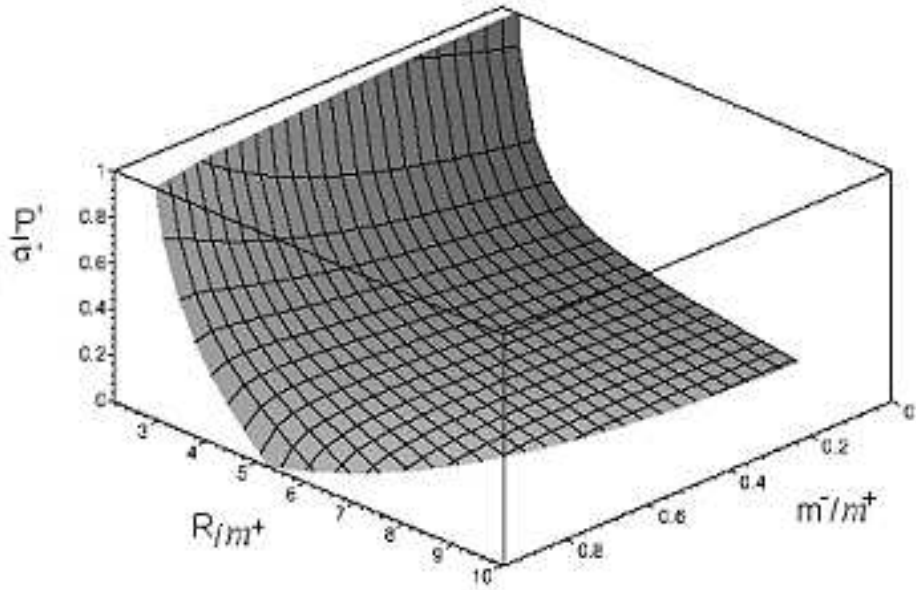}\\ \caption{Schwarzschild - de Sitter black
hole in a Schwarzschild background. Since $M$ and $\Upsilon$ are
$>0$ in this case, it follows that the region \textit{below} the
surface shown corresponds to stable equilibrium. The inclusion of
a positive cosmological constant inside a timelike shell weakens
the binding of the shell in the sense that $P$ becomes negative (a
surface \textit{tension}) to stablize the shell at sufficiently
large $R$. The result is quite unlike the case $\Lambda=0$ were
surface tensions are not required to stablize the shell at any
radius. Here $\Lambda m^{+2} \sim 10^{-3}$ well away from
degeneracy in all cases. } \label{fig3}
\end{figure}

\newpage

\begin{figure}
\psfig{file=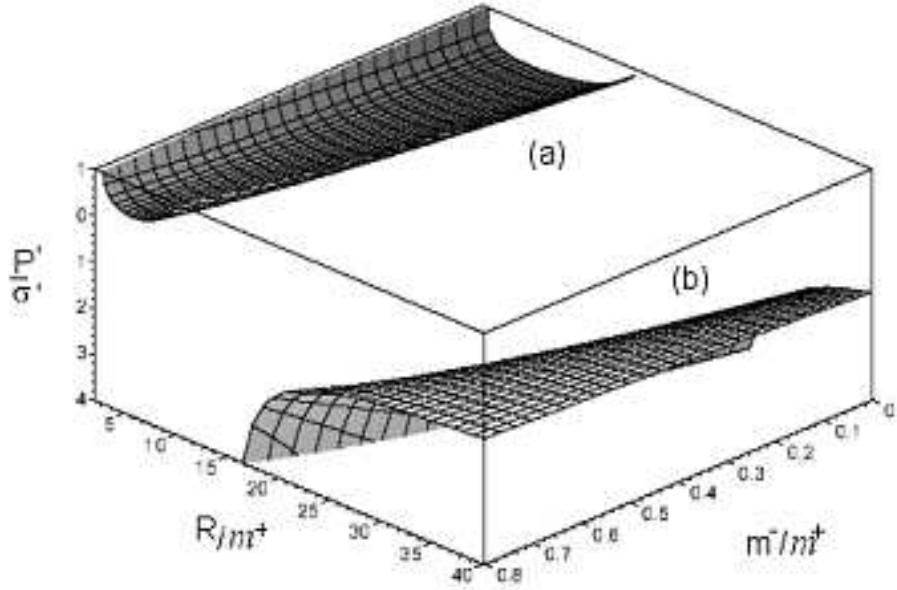}\\ \caption{Schwarzschild -anti de Sitter
black hole in a Schwarzschild background. This case is the same as
Figure 3 except that $\Lambda < 0$. Unlike the previous case
however, we do find an asymptote $\Upsilon=0$. Since $M>0$, it
follows from the form of $\Upsilon$ that the region \textit{below}
\textit{(a)} and \textit{above} \textit{(b)} are the regions of
stability. Here again $\Lambda m^{+2} \sim 10^{-3}$ well away from
degeneracy in all cases. } \label{fig4}
\end{figure}

\newpage

\begin{figure}
\psfig{file=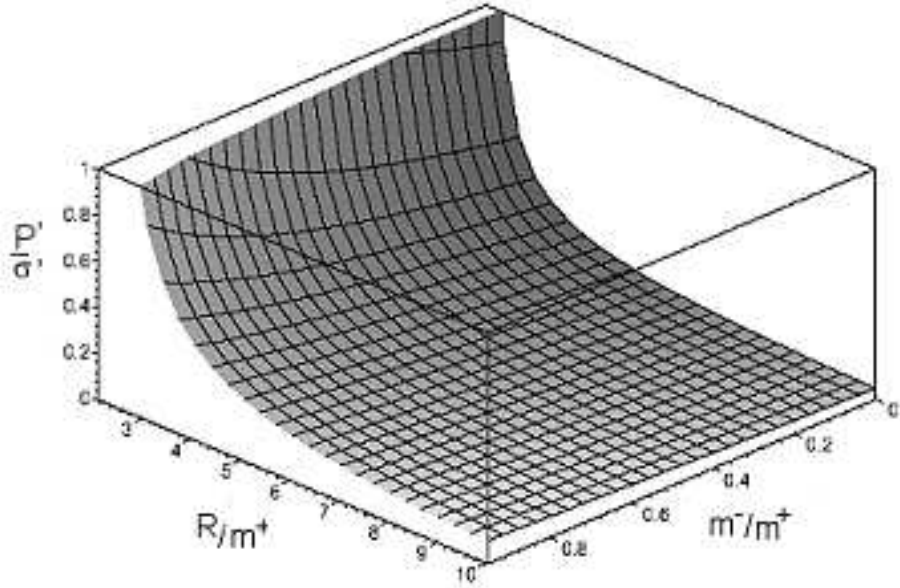}\\ \caption{Schwarzschild - de Sitter black
hole in a Schwarzschild - de Sitter background with $[\Lambda]=0$.
In this case one would expect that $\Lambda$ would not change the
qualitative features of equilibrium. This is exactly what we find
(see Figure 2). Here $\Lambda \texttt{m}^{+2} \sim 10^{-3}$ well
away from degeneracy in all cases. The diagram remains
qualitatively unchanged under a change in the sign of $\Lambda$ as
long as $[\Lambda]=0$. } \label{fig5}
\end{figure}

\newpage

\begin{figure}
\psfig{file=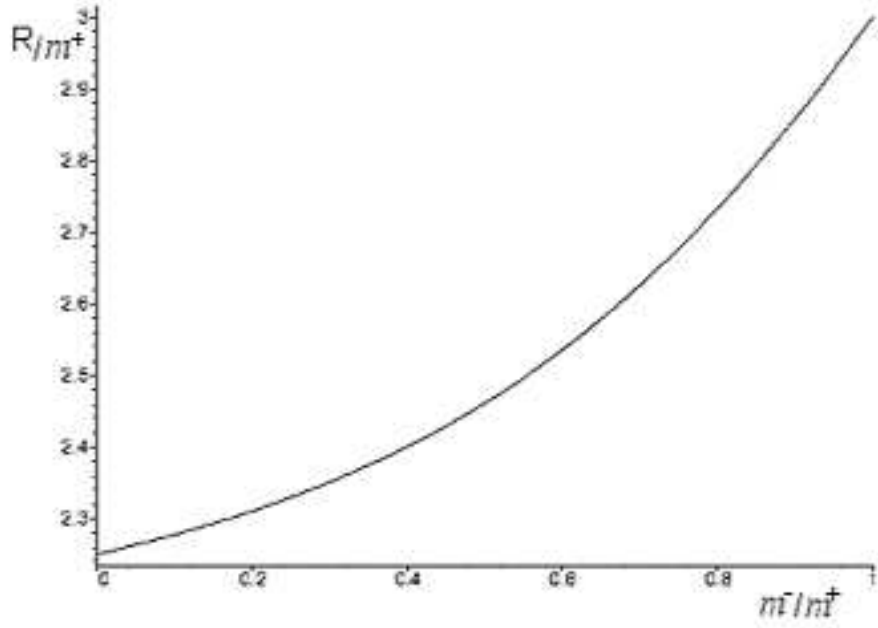}\\ \caption{The asymptote $\Upsilon=0$ for
the Schwarzschild wormhole (see Figure 1). The plot shows the
allowed roots to equation (\ref{upsilon}). } \label{fig6}
\end{figure}

\begin{figure}

\psfig{file=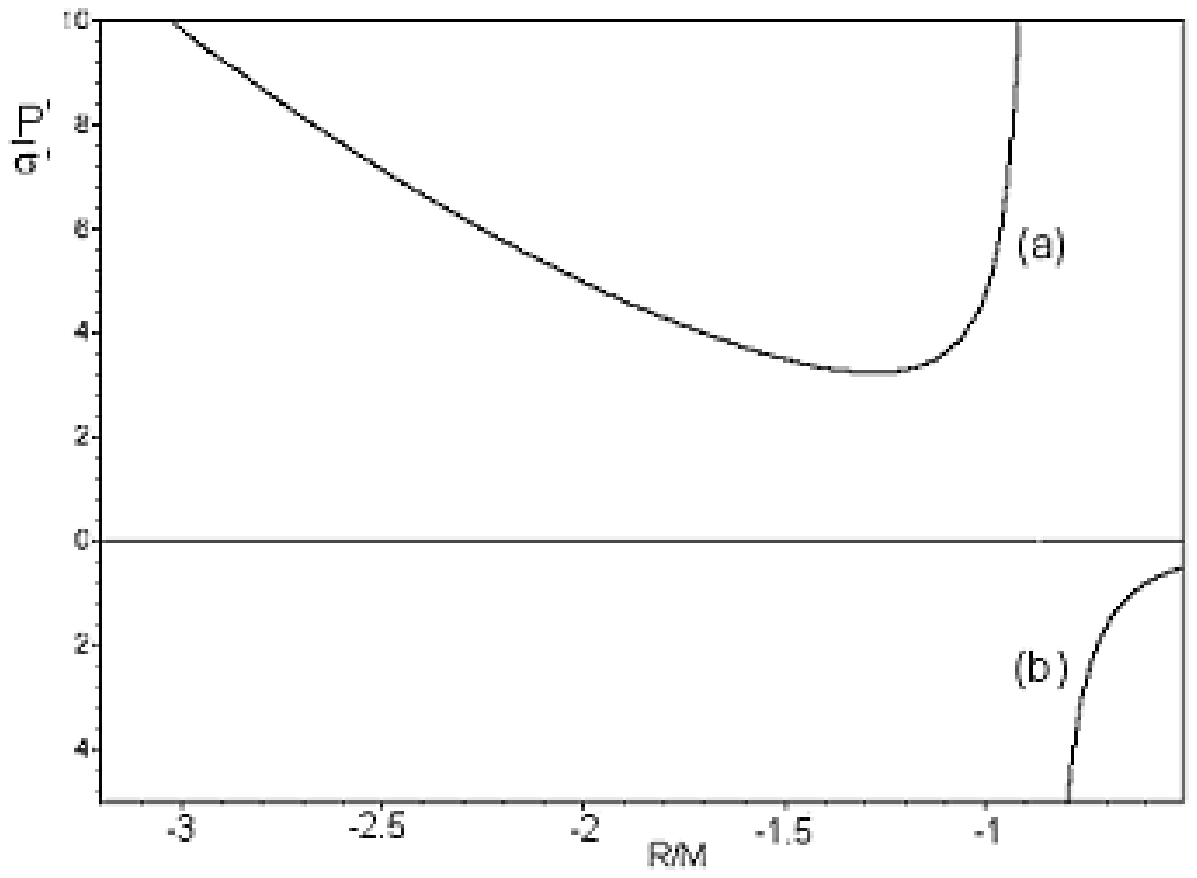}\\ \caption{Plot of stability regions versus
$R/M$ ($M$ is the intrinsic mass) for the case considered by
Poisson and Visser . The region \textit{above} the curve
\textit{(a)} and \textit{below} the curve \textit{(b)} correspond
to stable equilibrium. The position of the asymptote $\Upsilon=0$
is at $R/M=-\sqrt{3}/2$ corresponding to $R/m=3$.} \label{fig7}

\end{figure}


\begin{references}
\bibitem{domainwall} By ``domain wall" here we mean a hypersurface
with an intrinsics equation of state of the form $\sigma+P=0$,
where $\sigma$ is the surface density and $P$ is the surface
pressure. See \textit{e.g.} \cite{mus}.
\bibitem{notation} We write $\left[ X \right] \equiv X^+\mid_{\Sigma} - X^-\mid_{\Sigma}$
and $\overline{X} \equiv (X^+ \mid_\Sigma + X^- \mid_\Sigma)/2$
with $X^\pm\mid_{\Sigma}$ denoting the limiting values of $X$ on
$\Sigma$, and use geometrical units except in Appendix B. We use
the letter ``m" as follows: $M$ is the surface mass of $\Sigma$
defined by (\ref{surfacemass}), $m$ is the effective gravitational
mass of the enveloping spacetimes ``+" and ``-" (see \cite{mass}
and (\ref{bondi})), $[m] \equiv m^{+}-m^{-}$ and $\overline{m}
\equiv (m^{+}+m^{-})/2$. $\texttt{m}$ is not the effective
gravitational mass but is related to it for the cases considered
here by (\ref{mlambda}).
\end{references}
\end{document}